\documentclass[12pt,a4paper]{article}
\usepackage[margin=1in]{geometry}
\usepackage{setspace}
\usepackage{titlesec}
\usepackage{graphicx}
\usepackage{caption}
\usepackage{amsmath}
\usepackage{amsfonts}
\usepackage{cite}
\usepackage{authblk}
\usepackage{tabularx}
\usepackage{multirow}    
\usepackage{array}       
\usepackage{caption}

\titleformat{\section}{\bfseries\normalsize}{\thesection.}{1em}{}
\titleformat{\subsection}{\bfseries\normalsize}{\thesubsection.}{1em}{}

\begin{document}

\begin{center}
    {\LARGE\bfseries Learning from Limited Labels: Transductive Graph Label Propagation for Indian Music Analysis}

    \vspace{1em}
    {\large Parampreet Singh$^{1}$, Akshay Raina$^{2}$, Sayeedul Islam Sheikh$^{3}$, \\Vipul Arora$^{4}$}

    \vspace{0.5em}
    {\normalsize
    $^{1,2,4}$Department of Electrical Engineering, $^{3}$Department of Chemical Engineering\\ $^{1,2,3,4}$Indian Institution of Technology, Kanpur, India 
    $^{4}$Katholieke Universiteit Leuven\\
    Email: params21@iitk.ac.in$^{1}$, akshayy.rainaa@gmail.com$^{2}$, sayeedul21@iitk.ac.in$^{3}$, 	vipul.arora@kuleuven.be$^{4}$
    }
\end{center}

\vspace{1em}

\noindent\textbf{Abstract:}
\begin{quote}
Supervised machine learning frameworks rely on extensive labeled datasets for robust performance on real-world tasks. 
However, there is a lack of large annotated datasets in audio and music domains, as annotating such recordings is resource-intensive, laborious, and often require expert domain knowledge.
In this work, we explore the use of label propagation (LP), a graph-based semi-supervised learning technique, for automatically labeling the unlabeled set in an unsupervised manner. 
By constructing a similarity graph over audio embeddings, we propagate limited label information from a small annotated subset to a larger unlabeled corpus in a transductive, semi-supervised setting. 
We apply this method to two tasks in Indian Art Music (IAM): Raga identification and Instrument classification. 
For both these tasks, we integrate multiple public datasets along with additional recordings we acquire from Prasar Bharati\footnote{Prasar Bharati is India’s public broadcasting agency, comprising Doordarshan Television Network and All India Radio. It maintains an extensive
archive of Indian classical music recordings.} Archives to perform LP. 
Our experiments demonstrate that LP significantly reduces labeling overhead and produces higher-quality annotations compared to conventional baseline methods, including those based on pretrained inductive models. These results highlight the potential of graph-based semi-supervised learning to democratize data annotation and accelerate progress in music information retrieval.
\end{quote}

\vspace{0.5em}

\section{Introduction}
Most modern Machine Learning (ML) methods require extensive supervision for robust performance on real-world problems. However, the limited availability of labeled data and the resource-intensive cost of annotating instances pose significant bottlenecks to the scalability and deployment of ML solutions. This underscores the need for robust systems capable of assigning high-quality labels to raw examples.

This challenge is especially acute in audio or music domains, where labeling those recordings requires listening for long durations and maintaining precision for frame-level annotations. Although a wealth of audio/music datasets exist online\cite{gemmeke2017audio,fonseca2021fsd50k,piczak2015esc,IRMAS_juan_j_bosch_2018_1290750, salamon2014dataset,serra2014creating,sankalp_gulati_2016_7278511RRDdataset,bozkurt_2018_4301737saraga_dataset,emanuele2020tinysol,param_XAI,srinivasamurthy2015tablasolo,adithi_shankar_2024_14877279_saraga_audio_visual_dataset,sumitkumar2025RODdataset}, most suffer from scaling issues due to associated annotation and collection costs, which may lead to mislabeling~\cite{schmarje2022one}. This limitation corresponds to the limited size of such datasets. For instance, \cite{srinivasamurthy2015tablasolo} released the TablaSolo dataset of only 38 solo tabla (Indian percussion instrument) compositions. Similarly, \cite{piczak2015esc, salamon2014dataset} are datasets for sound event classification with only around 3 hours and 9 hours of recordings, respectively. 
There is a corpus of music databases~\cite{serra2014creating}, containing most datasets suffering from similar limitations.

It is therefore important to use systems capable of automatic annotation of audio or music recordings, while maintaining high label quality. 
Traditional ML systems that train on a labeled set and infer labels on the unlabeled set are ineffective solutions in such scenarios, as they require rich supervision~\cite{param_XAI,param_NCC_ontology,WIMAGA_confidence_Raga_ICASSP_2025}. One such solution is Label Propagation (LP), a semi-supervised technique that aims to learn from a sparsely labeled set and propagate labels onto a large unlabeled set using transductive learning. This approach has immense potential and has recently attracted substantial research interest~\cite{tian2023graphhop, pmlr-v139-cai21b, Tankasala2023}. It exploits two assumptions: (1) data points closer to each other are likely to have the same label; and (2) most data points on the same manifold should have the same label~\cite{zhou2003learning}. Unlabeled data points are labeled based on the similarity between their features and those of the labeled data points. Most LP algorithms are graph-based, where the edges between two nodes (instances) encode the affinity between them. This allows for effectively exploiting the underlying structure of the data to propagate labels through the graph, even with limited data. There have been notable works on LP for images~\cite{zhuѓ2002learning,zhou2003learning,pmlr-v139-cai21b,iscen2019label,zhu2023transductive}, but this has rarely been explored for audio/music datasets. 

LP offers an effective solution for large-scale metadata expansion, particularly in domains where labeled data is scarce but unlabeled data is abundant. Its task-agnostic nature allows it to be applied across a wide range of applications, making it ideal for annotating massive audio and music corpora sourced from platforms like YouTube, Spotify, or public archives. By leveraging the inherent structure in the data, LP provides a scalable and efficient alternative to manual annotation.

In this study, we employ a LP framework for two key tasks in Music Information Retrieval: Raga Identification~\cite{singh2025identificationclusteringunseenragas, param_NCC_ontology,param_XAI} and Instrument Recognition. We utilize multiple publicly available datasets in combination with a large corpus of unlabeled audio recordings from the Prasar Bharati Archives for carrying out LP across diverse musical content. 
Our results demonstrate that the proposed approach yields high-quality annotations, often outperforming several baselines, including fully supervised inductive learning approaches.

\section{Related Works}
The scarcity of labeled audio data has been a significant bottleneck in advancing machine learning applications in audio and music processing. 
Despite the growing interest in machine learning for audio and music processing, progress is often hindered by the limited size and scope of available labeled datasets. Many widely used resources, such as TinySOL~\cite{emanuele2020tinysol}, TablaSolo~\cite{srinivasamurthy2015tablasolo}, and IRMAS~\cite{bosch2012comparison}, are restricted by their size or the number of samples. Larger datasets like AudioSet~\cite{gemmeke2017audio} and FSD50K~\cite{fonseca2021fsd50k} offer broader coverage but often suffer from weak labeling and insufficient annotation detail for specialized music information retrieval tasks. 
For Indian classical music, datasets like the IAM Raga Recognition Dataset~\cite{sankalp_gulati_2016_7278511RRDdataset}, Saraga~\cite{bozkurt_2018_4301737saraga_dataset}, and PIM~\cite{param_XAI} provide valuable resources but are limited in size, require lots of manual labeling, and often focus on specific aspects like raga recognition without broader applicability.

These limitations in dataset size, diversity, and annotation quality highlight the need for approaches that can maximize the utility of limited labeled data. In this context, LP offers a promising solution by enabling the automatic extension of labels from a small annotated subset to much larger unlabeled collections, thus addressing a key bottleneck in music and audio machine learning research.

\textbf{Label Propagation} 
is a semi-supervised learning method that leverages the structure of the data manifold to propagate labels from a small set of labeled examples to a larger unlabeled set.
Early work by Zhu and Ghahramani~\cite{zhuѓ2002learning} introduced a LP algorithm using a fully connected graph where edge weights are determined by the Gaussian kernel of the Euclidean distance between data points. This method iteratively propagates labels while keeping the labeled data fixed.
Zhou et al.~\cite{zhou2003learning} extended this idea by incorporating both local and global consistency in the graph-based framework. They introduced a normalized graph Laplacian and formulated the LP as a closed-form solution, leading to efficient computation.
In recent years, Iscen et al.~\cite{iscen2019label} proposed a method that combines deep learning with LP. They use embeddings from a neural network to construct a sparse affinity matrix, which is then used in a diffusion process to propagate labels. This approach benefits from the representation power of deep networks and the structural information captured by the graph.

Our work leverages these advancements in label propagation and applies them to the music domain, specifically addressing the challenges of large-scale unlabeled datasets.

\section{Datasets}
\label{sec:dataset}
For both our tasks of Raga classification and Instrument classification, we leverage a combination of curated and publicly available datasets. Below, we describe the data sources and composition relevant to each task.

\subsection{Raga Classification}
For the Raga classification task, we use the PIM dataset introduced in \cite{param_XAI}, which consists of annotated audio recordings from Hindustani classical music performances. The dataset contains over 501 manually labeled audio files, totalling 23,365 audio chunks of 30 seconds, corresponding to 141 unique Ragas. It also includes additional metadata, including Raga, Tonic, Tala, Gharana, and performer annotations. Raga and Tonic labels have been manually annotated and verified for the dataset. It serves as the primary labeled source for our experiments. We apply LP to extend these annotations across a larger unlabeled corpus of audio recordings from Prasar Bharati archives.

\begin{table}[h]
\caption{Sampling from various public datasets for Instrument Recognition. The numbers represent the number of audio samples for each instrument. The instruments are: Accordian (Acc.), Cymbals (Cym), D.K. (Drum Kit), Guitar (Guit.), Organ, Piano, Tabla (Tab.), Trumpet (Trum.), Sitar (Sit.), Flute (Flu.), Violin (Vio.)}
\label{tab:1}
\centering
\resizebox{\textwidth}{!}{%
\begin{tabular}{|c|c|c|c|c|c|c|c|c|c|c|c|}
\hline
\textbf{Dataset} & \textbf{Acc.} & \textbf{Cym.} & \textbf{D.K.} & \textbf{Guit.} & \textbf{Organ} & \textbf{Piano} & \textbf{Tab.} & \textbf{Trum.} & \textbf{Sit.} & \textbf{Flu.} & \textbf{Vio.} \\ \hline
\textbf{FSD50K} & 99 & 835 & 351 & 2185 & 339 & 844 & 96 & 632 & - & - & - \\ \hline
\textbf{AudioSet} & - & - & - & - & - & - & 949 & - & 851 & 2697 & - \\ \hline
\textbf{IRMAS} & - & - & - & - & - & 721 & - & 577 & - & 451 & 580 \\ \hline
\textbf{TinySOL} & 689 & - & - & - & - & - & - & 96 & - & 118 & 284 \\ \hline
\textbf{Tabla Solo} & - & - & - & - & - & - & 38 & - & - & - & - \\ \hline
\end{tabular}%
}
\end{table}

\subsection{Instrument Recognition}

\begin{figure}[t]
  \centering
  \includegraphics[width=\textwidth]{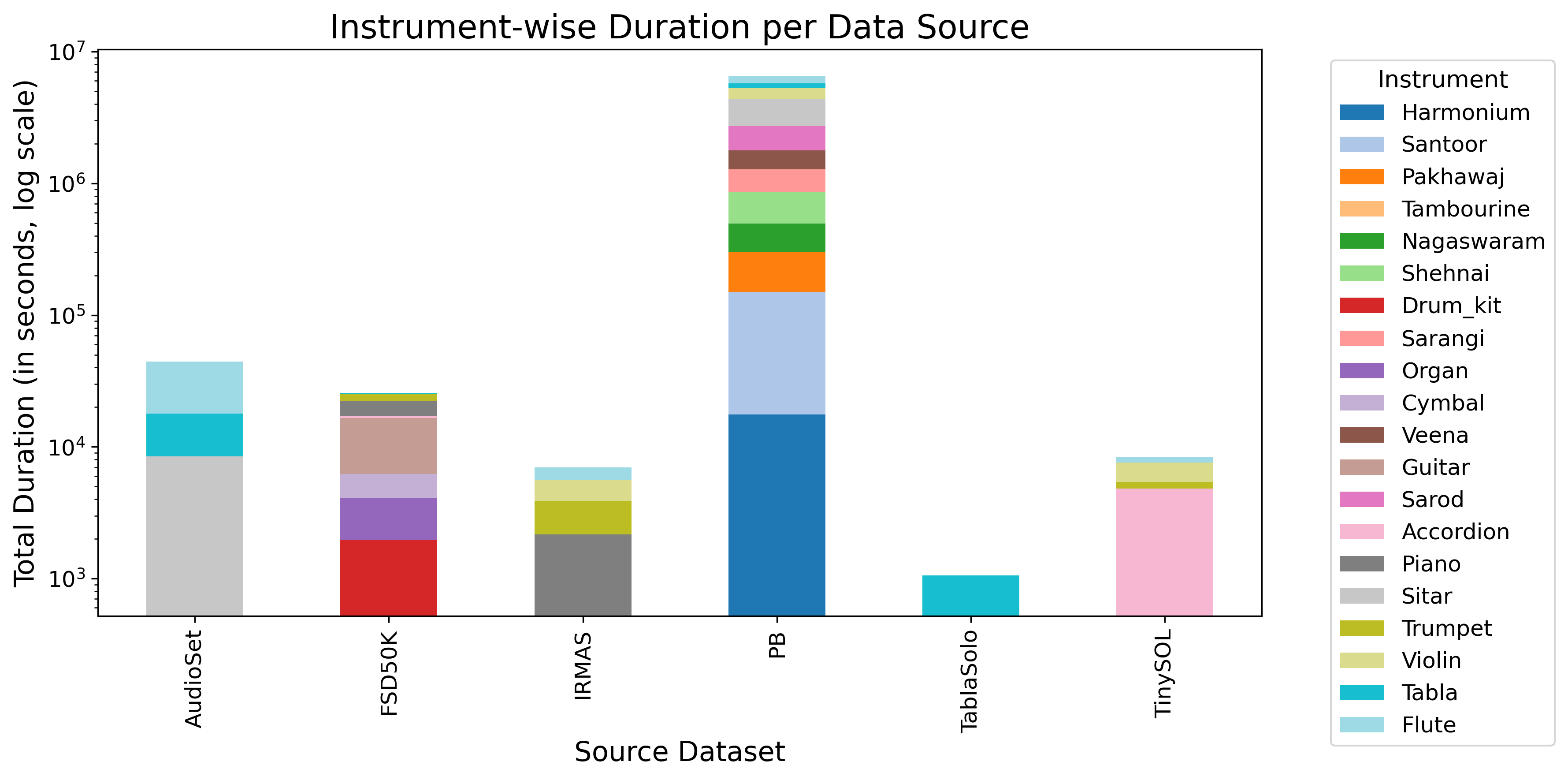}
  \caption{Source-wise duration of all instrument samples used from various datasets (in seconds). PB represents Prasar Bharati audios.}
  \label{fig:instrument_stats}
\end{figure}

For the Instrument Recognition task, we curate a dataset comprising recordings primarily sourced from the Prasar Bharati Archives, supplemented with samples from several publicly available datasets. 
The Prasar Bharati audios predominantly feature instruments which are commonly found in real-life Indian Classical Music performances, such as Sitar, Tabla, Veena, Pakhawaj, and Flute. 
A notable challenge in these audios is the imbalance across instrument classes—with approximately 65\% of the total duration concentrated in the top five most frequent classes. This imbalance can hinder the performance of machine learning models, especially during label propagation. To address this, we augment the Prasar Bharati audios by including class-specific samples from various open-source datasets, thereby improving class.
Specifically, we sample: TablaSolo~\cite{srinivasamurthy2015tablasolo} for Tabla recordings, FSD50K~\cite{fonseca2021fsd50k} for Guitar, Drum-kit, and Tabla, TinySOL\cite{emanuele2020tinysol} and IRMAS\cite{bosch2012comparison} for Flute and Violin, and AudioSet~\cite{gemmeke2017audio} for Flute and Drum-kit samples.
The complete source-wise distribution of instrument durations, including contributions from Prasar Bharati and external datasets, is shown in Figure~\ref{fig:instrument_stats}, while the number of samples acquired from other datasets is provided in Table~\ref{tab:1}.
The combined dataset includes a total of 20 instrument classes as shown in the legends in Figure~\ref{fig:instrument_stats}.

We divide the whole dataset into labeled and unlabeled sets and 
curate a gold set of 200 manually labeled and verified audio recordings out of the unlabeled set for evaluation purposes.

\section{Label Propagation}
\label{sec:4}

\begin{figure}[t]
  \centering
  \includegraphics[width=0.8\textwidth]{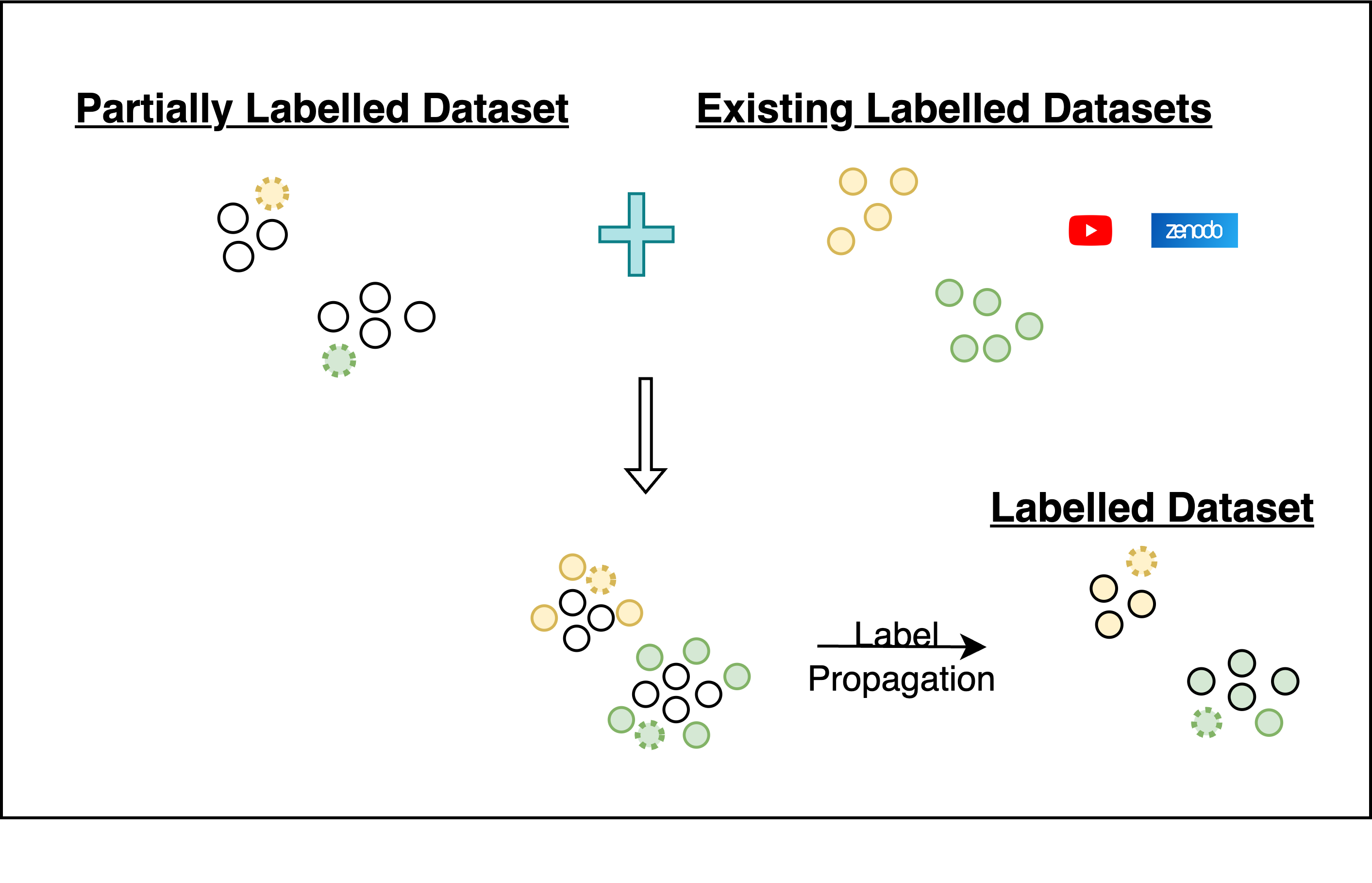}
  \caption{ Flowchart illustrating the Label Propagation process for automatic annotation of unlabeled audio samples. A small labeled dataset is combined with a larger set of partially labeled or unlabeled samples. Label Propagation is then applied using a similarity graph, enabling the transfer of label information from the labeled subset to the unlabeled data, resulting in the entire corpus being annotated in a transductive, semi-supervised manner.
  }
  \label{fig:Flow_chart}
\end{figure}

The scarcity of reliable labeled data necessitates the use of efficient transductive LP methods. We utilize pseudo-labels for unlabeled data to train a classifier and construct a graph by exploiting the embeddings obtained from the network~\cite{iscen2019label}. This is a two-fold method: (1) Train the network using the entire dataset (with pseudo labels for unlabeled data points), and (2) Construct a nearest-neighbor graph using the embeddings from the network.
This flowchart in Figure~\ref{fig:Flow_chart} illustrates the process of LP. Initially, a small set of audio samples with sparse labels (left) is combined with external labeled datasets sourced from platforms like YouTube and Zenodo (right). These datasets are merged to form a unified collection containing both labeled and unlabeled instances. Through the application of LP, label information from the annotated examples is extended to the unlabeled samples, resulting in a fully labeled dataset (bottom). This approach significantly reduces manual annotation effort while maximizing the utility of available data for downstream machine learning tasks.

\subsection{Classifier Training}

Let $X = \{x_1, x_2, \dots, x_l, x_{l+1}, \dots, x_n\}$ with $x_i \in \mathcal{X}$ be a collection of $n$ data points, where the first $l$ are labeled with class labels $y_i \in C$, and the remaining $u = n - l$ points are unlabeled. The label space is denoted by $C = \{1, 2, \dots, c\}$.
We employ a deep network consisting of a feature extractor $h_{\theta_1}: \mathcal{X} \rightarrow \mathbb{R}^d$ that maps each input $x_i$ to a $d$-dimensional embedding $z_i = h_{\theta_1}(x_i)$, and a classifier $g_{\theta_2}: \mathbb{R}^d \rightarrow \mathbb{R}^c$ that outputs class-wise confidence scores. The overall model is represented by $f_{\theta}(x) = g_{\theta_2}(h_{\theta_1}(x))$, where $\theta = (\theta_1, \theta_2)$.

The training objective involves multiple components:

\begin{align}
    L_s(X_l, Y_l; \theta) &= \sum_{i=1}^{l} l_s(f_{\theta}(x_i), y_i), \\
    L_p(X_u, \hat{Y}_u; \theta) &= \sum_{i=l+1}^{n} l_s(f_{\theta}(x_i), \hat{y}_i), \\
    L_u(X; \theta) &= \sum_{i=1}^{n} l_u(f_{\theta}(x_i), f_{\hat{\theta}}(\hat{x}_i)),
\end{align}

Here, $L_s$ is the supervised loss (typically cross-entropy) over the labeled data. $L_p$ is a pseudo-labeling loss computed using labels $\hat{y}_i$ predicted by LP. 
$L_u$ is the unsupervised loss term applied on $X_l \cup X_u$ to make the embeddings consistent across different transformations of an input.

\subsection{Transductive Label Propagation}

To generate labels $\hat{y}_i$ for the unlabeled examples, we use a transductive LP approach inspired by diffusion processes~\cite{zhou2003learning}. The core idea is to construct a similarity graph among all data points based on their embeddings and propagate known labels across the graph structure.

Let $W \in \mathbb{R}^{n \times n}$ be a symmetric adjacency matrix with zero diagonals, where each entry $w_{ij}$ captures the similarity between embeddings $z_i$ and $z_j$. This matrix is constructed using a symmetric $k$-nearest neighbor (k-NN) graph. We define $W = M + M^{T}$, where:

\begin{equation}
    m_{ij} = 
    \begin{cases}
        {[ (z_i^\top z_j)^\gamma ]_{+}}, & \text{if } i \neq j \text{ and } z_j \in \text{NN}_k(z_i), \\
        0, & \text{otherwise},
    \end{cases}
\end{equation}

where $\text{NN}_k(z_i)$ denotes the $k$ nearest neighbors of $z_i$ in embedding space, $\gamma$ is a sharpness hyperparameter, and $[\cdot]_+$ denotes just the positive part (ReLU function).

Next, we compute the symmetric normalized affinity matrix $S$ using:

\begin{equation}
    S = D^{-1/2} W D^{-1/2},
\end{equation}

where $D = \mathrm{diag}(W \mathbf{1}_n)$ is the degree matrix and $\mathbf{1}_n$ is an all-ones vector of length $n$.

We now construct a label matrix $Y$ of shape $n \times c$ such that $Y_{ij} = 1$ if $x_i$ is labeled and its class is $j$, and $Y_{ij} = 0$ otherwise. Thus, labeled examples are one-hot encoded, and unlabeled rows are all zeros.
LP computes soft labels over all nodes via the closed-form solution:

\begin{equation}
    P = (I - \alpha S)^{-1} Y,
\end{equation}

where $P \in \mathbb{R}^{n \times c}$ contains the propagated label distributions, and $\alpha \in [0, 1)$ controls the strength of label diffusion. The rows of $P$ can be interpreted as class probability scores.
Finally, we assign pseudo-labels using:

\begin{equation}
    \hat{y}_i = \arg\max_j P_{ij}.
\end{equation}

This assigns to each example $x_i$ the class with the highest propagated score.
Here, it is noteworthy that computing the matrix inverse $(I - \alpha S)^{-1}$ is computationally infeasible for large $n$ because it is not sparse. Instead, following~\cite{iscen2019label}, we solve the linear system:

\begin{equation}
    (I - \alpha S) Z = Y
\end{equation}

using the conjugate gradient method, which yields $Z \approx P$. 
We iteratively propagate these pseudo-labels while jointly optimizing the loss functions described earlier, ensuring that the overall loss continues to decrease across iterations. Once convergence is achieved, the final pseudo-labels obtained from the diffusion process are used as predictions for evaluation.

\begin{table}[t]
\centering
\hspace{27pt}
\begin{tabular}{|c|c|c|c|}
\hline
                & \textbf{Precision} & \textbf{Recall} & \textbf{F1} \\ \hline
\textbf{Speech} & 1.0                & 0.93            & 0.97        \\ \hline
\textbf{Music}  & 1.0                & 0.98            & 0.99        \\ \hline
\end{tabular}
\caption{
Performance of the PANNs~\cite{kong2020panns} model on the labeled portion of the PIM dataset for the Speech vs. Music classification task. The model is evaluated on 501 Hindustani classical music recordings annotated with speech and music segments. Metrics are computed at the 30-second chunk level after excluding ambiguous segments containing both speech and music.
}
\label{tab:panns_speech}
\end{table}
\section{Experiments}

We utilized the LP scheme discussed in Section~\ref{sec:4} to expand the metadata for Music Instrument Recognition and Raga Identification tasks. 
First, we train a Deep Neural Network in a fully supervised setup on 80\% of the labeled set $X_l$. We then infer labels from this trained network for the remaining 20\% labeled recordings and all unlabeled recordings $X_u$. To assess the performance on the unlabeled set, we manually annotate and verify a subset of recordings from $X_u$. Finally, we report the accuracy obtained on the 20\% held-out set (labeled) and the manually annotated recordings from the unlabeled set.
We now explain experimental details for both tasks in detail.


\subsection{Music Instrument Recognition}
For the instrument detection task, we pre-process the audio recordings by first discarding all files shorter than 1 second. The remaining files are segmented into 5-second chunks, with each chunk inheriting the instrument label of its source recording. Mel-spectrograms are extracted from each chunk using a window size of 1024, hop length of 512, and 64 mel bins. These serve as input features to the network.

As a baseline, we use a modified ResNet-18\cite{he2016deep} trained in a fully supervised fashion. For our proposed approach, we apply LP as described in Section\ref{sec:4}, leveraging a small labeled subset and a larger pool of unlabeled examples.
Both models are trained for a maximum of 50 epochs using the Adam optimizer with Stochastic Gradient Descent. For evaluation, we reserve a manually annotated gold test set of 200 test audio recordings, and accuracy is computed on the held-out test set to assess performance.

\subsection{Raga Identification}
For the Raga Identification task, we work with a total of 61,705 audio chunks, 30 seconds each, out of which 13,075 are labeled and taken from PIM~\cite{param_XAI} dataset and span 41 known Raga classes, along with a 42nd \textit{Others} class. 
The remaining 48630 unlabeled recordings are sourced from the Prasar Bharati Archives. 
To construct the evaluation set, for each of the 42 Raga classes, we select at least one full audio file out of the labeled set based on their representation. These selected recordings are then split into 30-second chunks, and all resulting chunks are included in the evaluation set, resulting in a total of 3,210 evaluation chunks. During LP training, these chunks are merged with the unlabeled set by discarding their true labels, creating a transductive learning setting.

\begin{figure}[t]
  \centering
  \includegraphics[width=\textwidth]{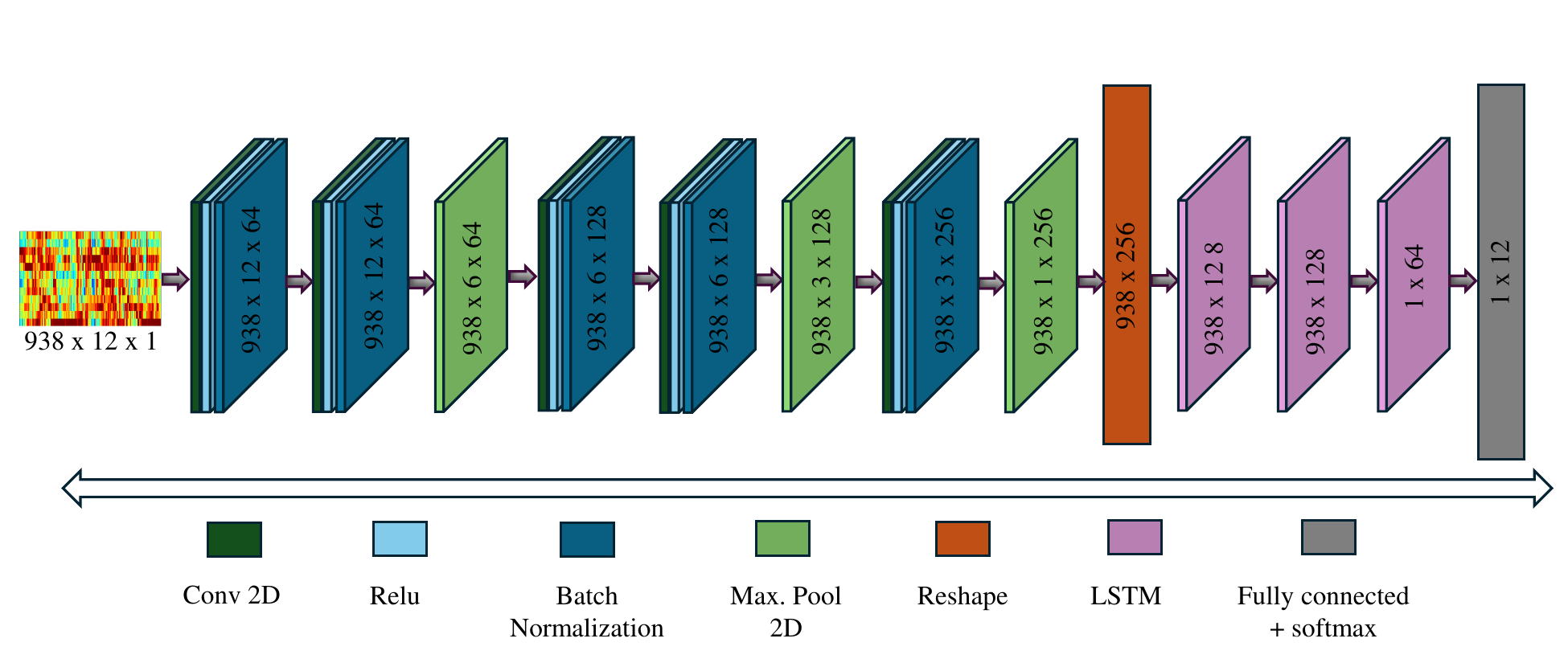}
  \caption{Model architecture used for the Raga Identification task. The model consists of convolutional layers followed by LSTM layers to capture temporal dependencies in the audio data.}
  \label{fig:Raga_model}
\end{figure}
To filter out non-musical (speech) segments, we first annotate the 501 recordings present in the PIM\cite{param_XAI} dataset with speech and music timings and evaluate the performance of the PANNs model~\cite{kong2020panns} for automatic segmentation.
The PANNs model is tested on 501 Hindustani classical music files, comprising 23,224 audio chunks of 30 seconds each. Chunks containing overlapping speech and music are excluded due to labeling ambiguity. Predictions of PANNs are evaluated at the chunk level. For cases where the model’s top prediction is not \textit{Music}, we consider it music if one of the top predicted classes includes a relevant musical tag (e.g., tabla, sitar) and \textit{Music} is the second-highest class.
Out of 23,224 chunks, 21,691 are classified as music, 1,453 as speech, and 80 as ambiguous. As shown in Table~\ref{tab:panns_speech}, the PANNs model shows robust and reliable performance for this task, and can be used for labeling all other audio files from PB recordings.

We train our model for a 42-class classification task with 41 known Ragas (Having the most number of audio files in the labeled dataset) and an additional \textit{Others} class, which includes all other remaining Ragas and speech segments extracted from the audio files themselves using the PANNs model. For tonic normalisation, we use the tonic values provided in \cite{param_XAI}, and for the remaining audios, we use
the \textit{CompIAM} package~\cite{compiam_mtg_2023} for computing tonic values, which is known to be very useful for the task, as explained in \cite{param_XAI}.

We employ a CNN-LSTM model architecture for our classification task, as shown in Fig.~\ref{fig:Raga_model}. Initially, we train it in a fully supervised manner using the given labels and evaluate it on the test set extracted from the labeled data. Next, we use the same architecture, pre-train it for just 10 epochs, and then use it for feature extraction and deploy the LP method to train it using both the labeled and unlabeled examples. For evaluation, we assess our models at both the audio chunk level and the audio file level. At the chunk level, we measure the accuracy of classification for each audio chunk, while at the audio file level, we determine the majority vote among all chunks sourced from the same audio file for class prediction.

\section{Results and Discussion}

\paragraph{Instrument Recognition.} 
Table~\ref{tab:instrument_recog} shows the performance comparison between the supervised baseline and the LP method. Accuracy is reported on two sets: Acc.1 refers to a held-out portion of the labeled set ($X_l$), and Acc.2 refers to a manually annotated subset of the originally unlabeled set ($X_u$). The LP method achieves 84.6\% on Acc.1 and 91.7\% on Acc.2, compared to 48.3\% and 63.2\% respectively for the supervised baseline. This demonstrates that the propagated labels are of high quality and the model predictions after LP are quite trustworthy, even on data not seen during training.

\begin{table}[t]
    \centering
    \caption{Performance Analysis of Label Propagation with Baseline on Instrument Recognition. Acc.1 and Acc.2 correspond to the accuracies on subsets of $X_l$ and $X_u$, respectively.}
    \label{tab:instrument_recog}
    \begin{tabular}{|c|c|c|}
        \hline
        \textbf{Method} & \textbf{Acc.1 (\%)} & \textbf{Acc.2 (\%)} \\ \hline
        \textbf{Baseline} & 48.3 & 63.2 \\ \hline
        \textbf{Label Propagation} & \textbf{84.6} & \textbf{91.7} \\ \hline
    \end{tabular}
\end{table}

\paragraph{Raga Classification.} 
Table~\ref{tab:Raga_class} presents the performance metrics for Raga classification, evaluated at both the chunk level (CL) and file level (FL. The supervised model achieves an F1-score of 0.60 (CL) and 0.77 (FL). With the application of LP, these scores improve to 0.62 (CL) and 0.82 (FL). While chunk-level gains are modest, the improvement at the file level is significant (5\% absolute increase in F1-score). This is crucial for our application, which involves labeling full-length music recordings.
Although the model is trained on a fixed set of Raga classes, it is designed to handle unseen or ambiguous cases by assigning such instances to a generic \textit{Others} category.

\begin{table}[t]
    \centering
    \caption{Performance Comparison for Raga Classification task. CL: Chunk Level, FL: File Level.}
    \label{tab:Raga_class}
    \begin{tabular}{|c|c|c|c|}
        \hline
        \textbf{Method} & \textbf{Precision} & \textbf{Recall} & \textbf{F1 Score} \\ \hline
        \textbf{Supervised (CL)} & 0.63 & 0.57 & 0.60 \\ \hline
        \textbf{Supervised (FL)} & 0.76 & 0.80 & 0.77 \\ \hline
        \textbf{Label Propagation (CL)} & 0.65 & 0.59 & \textbf{0.62} \\ \hline
        \textbf{Label Propagation (FL)} & 0.83 & 0.82 & \textbf{0.82} \\ \hline
    \end{tabular}
\end{table}

These results reinforce the effectiveness of the LP framework in generating meaningful pseudo-labels for the two IAM tasks. The technique can be easily implemented to any downstream MIR classification tasks. 
These performance improvements after LP have practical use cases such as automatic cataloging or metadata generation in music archives.
Overall, the LP method consistently enhances performance across tasks and demonstrates its potential as a reliable and scalable labeling strategy for large, partially labeled music datasets.

\section{Conclusions and Future Work}

In this paper, we explore the use of label propagation, a graph-based semi-supervised learning technique, to address the challenge of limited labeled data for Music Information Retrieval tasks. 
We focus on two key tasks within the domain of Indian Art Music (IAM): Raga identification and Instrument classification. 
By constructing a similarity graph over audio segments, we exploit the inherent structure in the feature space
to propagate labels from a small labeled subset to a much larger unlabeled corpus in a transductive, semi-supervised manner.
Our results demonstrate that label propagation is an effective alternative to traditional supervised approaches, especially in domains like IAM research, where expert annotations are expensive, time-consuming, and require deep domain expertise. 
The method yields high-quality pseudo-labels and enables scalable annotation, making it well-suited for settings where labeled data is scarce but unlabeled data is abundant.

This study highlights the broader utility of graph-based semi-supervised learning for developing robust MIR systems with minimal manual labeling. Future directions include exploring adaptive or dynamic graph construction, incorporating temporal and structural musical features into the propagation process, and extending the framework to support open-set recognition, where previously unseen labels are automatically detected and modeled rather than being grouped under a generic “Others” category. Such extensions would make the system more realistic and applicable in real-world scenarios where new categories continuously emerge.

\section*{Acknowledgments}

We would like to thank Prasar Bharati (PB), India for providing the audio recordings and metadata. We also acknowledge the contributions of Renu Chavan and Kajal Heer for their efforts in data annotation and validation.

\bibliographystyle{IEEEtran}
\bibliography{TASLP.bib}
\end{document}